\documentclass[journal]{IEEEtran}
\usepackage[utf8]{inputenc}

\usepackage{graphicx}
\usepackage{dblfloatfix} 
\usepackage{color,soul}
\usepackage{mathtools}
\usepackage{amsmath, amssymb, amsthm, bm}  
\usepackage{todonotes}
\usepackage{booktabs}
\usepackage{cite}
\usepackage{enumerate}
\usepackage{units}
\usepackage{url}
\usepackage[algoruled]{algorithm2e}

\theoremstyle{definition} % bold title, normal text

\newtheorem{definition}{Definition}
\theoremstyle{definition} % bold title, normal text

\theoremstyle{remark} % italic title, normal text
\newtheorem{remark}{Remark}

\newcommand{\soc}[1]{\left\lVert#1\right\rVert_2} 

\makeatletter
\newcommand{\removelatexerror}{\let\@latex@error\@gobble}
\makeatother

\usepackage{tikz}
\usetikzlibrary{positioning}
\usepackage{pgfplots}
\usepgfplotslibrary{groupplots}
\pgfplotsset{compat=1.13}
\pgfplotsset{scaled y ticks=false}

\usepackage[noabbrev,capitalize]{cleveref}

\crefname{equation}{}{}
\Crefname{equation}{}{}

\newcommand{\set}[1]{\mathcal{#1}} % for caligraphed set-symbols
\DeclareMathOperator{\PTDF}{PTDF}
\DeclareMathOperator{\LODF}{LODF}

\newenvironment{ldescription}[1]
  {\begin{list}{}%
   {\renewcommand\makelabel[1]{##1\hfill}%
   \settowidth\labelwidth{\makelabel{#1}}%
   \setlength\leftmargin{\labelwidth}
   \addtolength\leftmargin{\labelsep}}}
  {\end{list}}

\title{Fast Security-Constrained Optimal Power Flow through Low-Impact and Redundancy Screening}

\author{Richard Weinhold and Robert Mieth
\thanks{The authors gratefully acknowledge the support by the German Federal Ministry for Economic Affairs and Energy (BMWi) in the project “Long-term Planning and Short-term Optimization of the German Electricity System Within the European Context” (LKD-EU, 03ET4028A). The work of R. Mieth was supported by the Reiner Lemoine-Foundation. \textit{(R. Weinhold and R. Mieth contributed equally to this work.)} \textit{(Corresponding author: R. Weinhold.)}}
\thanks{R. Weinhold is with the Fakultät VII Wirtschaft und Management, Technische Universität Berlin, 10623 Berlin, Germany (e-mail: riw@wip.tu-berlin.de).}
\thanks{R. Mieth is with the Department of Electrical and Computer Engineering, Tandon School of Engineering, New York University, New York, NY 10012 USA, and also with the Fakultät IV Elektrotechnik und Informatik, Technische Universität Berlin, 10587 Berlin, Germany (e-mail: robert.mieth@nyu.edu).}
}

\begin{document}
\bstctlcite{IEEE:BSTcontrol} % To activate IEEEtran.bst controls in .bib file

\maketitle

\begin{abstract}
Determining contingency aware dispatch decisions by solving a security-constrained optimal power flow (SCOPF) is challenging for real-world power systems, as the high problem dimensionality often leads to impractical computational requirements.
This problem becomes more severe when the SCOPF has to be solved not only for a single instance, but for multiple periods, e.g. in the context of electricity market analyses. 
This paper proposes an algorithm that identifies the minimal set of constraints that exactly define the space of feasible nodal injections for a given network and contingency scenarios.
By internalizing the technical limits of the nodal injections and enforcing a minimal worst-case impact of contingencies to line flows, computational effort can be further improved.
The case study applies and analyzes the methods on the IEEE 118 and A\&M 2000 bus systems, as well as the German and European transmission systems.
In all tested cases the proposed algorithm identifies at least \unit[95]{\%} of the network and security constraints as redundant, leading to significant SCOPF solve time reductions.
Scalability and practical implementation are explicitly discussed.
The code and input data of the case study is published supplementary to the paper under an open-source license. 
\end{abstract}

% \begin{IEEEkeywords}
%     
% \end{IEEEkeywords}{}

\section*{Nomenclature}

\noindent 
\textit{A. Sets}
\begin{ldescription}{$xxxxxx$}
\item [$\emptyset$] The empty set
\item [$\set{C}$] Set of contingency scenarios with $C=|\set{C}|$
\item [$\set{F}(B,\overline{f})$] Feasible region of system $(B,\overline{f})$
\item [$\set{G}$] Set of generators with $G=|\set{G}|$
\item [$\set{I}, \set{J}$] Set of indices
\item [$\set{L}$] Set of lines/edges with $L=|\set{L}|$
\item [$\set{L}_c$] Set of failed lines at contingency scenario $c$, $\set{L}_c \subseteq \set{L}$
\item [$\set{L}_n$] Set of lines connected to node $n$, $\set{L}_n \subseteq \set{L}$
\item [$\set{N}$] Set of nodes with $N=|\set{N}|$
\item [$\set{T}$] Set of time steps indexed with $T=|\set{T}|$
\end{ldescription}

\noindent
\textit{B. Parameters and Variables}
\begin{ldescription}{$xxxxxx$}
\item [$d_t$] Active power demand at $t$ indexed by $d_{t,n}, n\in\set{N}$
\item [$f_t$] Active power flow at $t$ indexed by $f_{t,l}, l\in\set{L}$
\item [$\overline{f}$] Maximum line capacity indexed by $\overline{f}_l, l\in\set{L}$
\item [$g_t$] Active power generation at $t$ indexed by $g_{t,i}, i \in\set{G}$
\item [$\underline{g}_t$] Lower generation limit at $t$ indexed by $\underline{g}_{t,i}, i \in\set{G}$
\item [$\overline{g}_t$] Upper generation limit at $t$ indexed by $\overline{g}_{t,i}, i \in\set{G}$
\item [$x_t$] Nodal injection at $t$ indexed by $x_{t,n}, n\in\set{N}$
\item [$\check{x}, \hat{x}$] Asymmetric bounds on nodal injection
\item [$\underline{x}, \overline{x}$] Symmetric bounds on nodal injection
\item [$B$] Generalized power transfer distribution matrix
\item [$I$] Identity matrix of appropriate dimensions
\item [$\LODF_{lc}$] Line outage distribution factor for line $l$ under $c$
\item [$M$] Mapping of generators to nodes
\item [$\eta$] Impact screening margin 
\end{ldescription}

\noindent
\textit{C. Operators}
\begin{ldescription}{$xxxxxx$}
\item [$|\set{X}|$] Cardinality of set $\set{X}$
\item [$\set{X}^\circ$] Interior of set $\set{X}$
\item [$X^{\!\top}$] Transpose of matrix $X$
\item [$X_i$] Row vector equal to the $i$-th row of $X$
\item [$X_{ij}$] $j$-th entry in the $i$-th row of $X$
\end{ldescription}

\section{Introduction}

\IEEEPARstart{P}{ower} flow physics and transmission limits constrain electricity market transactions.
With increasing uncertainty, mainly driven by the proliferation of intermittent renewable generation, a precise calculation of securely available transmission capacity can improve market efficiency and system reliability, \cite{powerfactsEurope2019}.
For example, in 2015 the transmission system operators (TSOs) of central western Europe (CWE) inaugurated flow based market coupling (FBMC) to manage cross-border electricity trading on the shared transmission infrastructure, ``\textit{bringing commercial transactions closer to the physical reality}'', \cite{amprion_2018}.
As a result, FBMC introduces operative security considerations into the market clearing process by identifying critical network elements and outage scenarios (contingencies), so called critical branches under critical outages (CBCOs), \cite{van2016flow}.
The identification of these CBCOs requires power flow optimization with contingency scenarios and is therefore closely related to solving a security-constrained optimal power flow (SCOPF).
As FBMC represents a significant part of the market clearing in Europe, \cite{schonheit2020impact}, studies on its interconnected markets need some means of accommodating a representation of the security constrained transmission infrastructure.
To enable multiperiod market simulations that internalize physical network constraints with contingency scenarios, this paper proposes a method that identifies the minimal set of constraints that defines the solution space spanned by the transmission and contingency constraints. 
Once acquired, this set can be used to significantly reduce the computational effort of solving SCOPF problems on this network. 

\subsection{Related Literature}
 
Since its introduction in \cite{alsac1974optimal}, SCOPF and its solution has been studied extensively.
Solving a full SCOPF problem in practice is typically obstructed by the dimensions of the resulting numerical problem and its computational complexity, \cite{capitanescu2011state}.
However, it is well known that only a limited subset of credible contingencies will eventually be active at the optimal solution, \cite{bouffard2005umbrella}.
Leveraging the fact that a candidate solution can easily be checked for feasibility, standard state-of-the-art approaches rely on iteratively adding contingency scenarios to a reduced base problem.
This method was proposed in \cite{alsac1974optimal,stott1978power} and has since been improved and extended, e.g. by more efficient constraint selection methods based on line loading, \cite{wood1996power}, impact bounds, \cite{brandwajn1988efficient}, or a ranking of corrective actions, \cite{fliscounakis2013contingency}. 
Further extensions towards corrective control actions have been proposed, e.g. in  \cite{capitanescu2008new}, and \cite{karangelos2019iterative} includes the risk of failure of these actions.
To decrease solution time, decomposition techniques based on Bender's Decomposition, \cite{li2008decomposed,phan2013some,dvorkin2016optimizing,velloso2019exact}, or decentralized optimization based on the Alternating Direction Method of Multipliers (ADMM), \cite{phan2013some,chakrabarti2014security,chakrabarti2020look}, have been proposed.
Departing from the iterative solution approach, recent work in \cite{thams2017data,halilbavsic2018data} use data-driven decision making trees to map contingencies to conditional line transfer capacities that ensure secure and stable post-contingency operation. 

The methods in \cite{stott1978power,wood1996power,brandwajn1988efficient,fliscounakis2013contingency,capitanescu2008new,karangelos2019iterative,li2008decomposed,dvorkin2016optimizing,velloso2019exact,phan2013some,chakrabarti2014security,chakrabarti2020look,thams2017data,halilbavsic2018data} are concerned with the solution of a specific instance of the SCOPF problem.
As such, they can be feasible in the context of day-to-day operational computations, but might be too complex for simulations that require solving an SCOPF for multiple time steps, even under the DC power flow assumption. 
Alternatively, the SCOPF can be simplified by identifying constraints that are never active (redundant). 
Respective methods haven been proposed in the context of network-constrained unit commitment without contingencies, e.g. in \cite{zhai2010fast,hua2013eliminating,roald2019implied,pineda2019data}.
In the context of a mixed-integer security constrained unit commitment problem, \cite{madani2016constraint} define bounds on the decision variables and then identify constraints that are redundant given these bounds.
A two step method combining a data-driven pre-screening and an iterative reintroduction of previously discarded constraints was proposed in \cite{zhang2019data}.

Our approach in this paper is closest related to the notion of ``umbrella contingencies'', which have been introduced in \cite{bouffard2005umbrella}. 
This contingency (sub)set contains the most restrictive outages that cover for all other possible outages implicitly and is independent from case-specific objective functions and uncertain parameters, such as load profiles or renewable generation.
While computing the SCOPF with this subset of contingencies has been shown to significantly reduce its solve time, identifying this set, on the other hand, is itself obstructed by impractically high computational effort.
To improve umbrella constraint discovery, network partitioning to enable parallel computation, \cite{ardakani2013identification}, and approximate pre-processing, \cite{ardakani2014acceleration}, has been proposed.
However, these approaches require an additional layer of implementation and the results are sensitive to the partitioning method.
In \cite{ardakani2018prediction} neural networks are proposed to predict umbrella constraints if system conditions change, but this requires previous identification of training sets.

\subsection{Contributions}

Similar to \cite{ardakani2014acceleration,ardakani2013identification}, this paper proposes a method to identify a minimal set of constraints that exactly represent the space of feasible nodal injections defined by the transmission system and its contingency scenarios.
Relative to \cite{ardakani2014acceleration,ardakani2013identification} we improve the discovery of this minimal set by leveraging a geometric algorithm based on \cite{clarkson1994more} that scales better and avoids additional preprocessing.
Also, no pregenerated or historical samples are required. 
The resulting minimal set directly relates to the CBCOs, as it collects constraints of transmission lines that are critical under some critical outages. 

Additionally, we demonstrate two methods that further reduce the number of necessary constraints, such that the time needed to identify these constraints and the solve time of the subsequent SCOPF is further decreased.
First, we propose to only include constraints that are non-redundant under the condition that the nodal injection at each node does not exceed predefined technical limits. 
In geometric terms, these technical nodal injection limits add additional cuts to the SCOPF solution space and generally remove more than one of the CBCO-related constraints.
Second, we show how a large number of constraints can be ignored, if the impact of contingencies on line flows is below a certain threshold and relate this approach to the common operational practice of security margins.
The threshold choice is discussed in terms on its impact on the SCOPF solution.

The proposed methods have been implemented in an open-source framework, which is tailored to enable comprehensive analyses of the European power system and markets with FBMC market clearing, \cite{pomato}.
To showcase the performance of the proposed methods, we conduct numerical experiments on the illustrative IEEE 118 bus system, on two real-world data-sets of the German and European power system, and on the A\&M synthetic 2000 bus network.
Finally, we discuss scalability and practical implementation.

\section{Problem Formulation}

In this paper we consider a preventive SCOPF problem on multiple time steps $t\in\set{T}$. 
As common for this type of analyses we leverage the DC power flow approximation, \cite{capitanescu2011state,ardakani2013identification}, to derive a linear relationship between nodal active power injections, contingencies and line flows.

\subsection{Power Flow Preliminaries}
\label{ssec:prelim_power_flow}

The physical network is represented by the set of nodes $\set{N}$ with $N = |\set{N}|$, the set of generators $\set{G}$ with $G = |\set{G}|$ and the set of lines $\set{L}$ with $L = |\set{L}|$.
Vector $g_t$ indexed by $g_{t,i} \geq 0$ denotes the active power generation of each generator $i$ and vector $d_t$ indexed by $d_{t,n} \leq 0$ denotes the aggregated active demand at each node $n \in \set{N}$ at time $t$.
At every time $t$ the vector of nodal injections is given by 
\begin{align}
    x_t = d_t + Mg_t \label{eq:def_nodal}   
\end{align}
indexed by $x_{t,n}$ where $M$ is a mapping of generators to nodes. 
Upper and lower generation limits are given by $\overline{g}_t$ and $\underline{g}_t$ indexed by $\overline{g}_{t,i}$ and $\underline{g}_{t,i}$ respectively. 
Each line $l \in \set{L}$ is a directed connection with arbitrary but fixed orientation between one sending node $s$ and one receiving node $r$.
At each time $t$ positive flow $f_{t,l} \geq 0$ indicates active power flow from $s$ to $r$ and negative flow $f_{t,l} \leq 0$ indicates active power flow from $r$ to $s$ over line $l$.
For all $l \in \set{L}$ the power flows are collected in the vector $f_t$ indexed by $f_{t,l}$ and the line capacities are given by vector $\overline{f}$ indexed by $\overline{f}_l$.
The physical power flow equations are approximated by power transfer distribution factors (PTDFs) where the PTDF matrix $B^0 \in \mathbb{R}^{L\times N}$ is a linear mapping of nodal injections $x_t$ to power flows $f_t$ such that:
\begin{align}
    f_t = B^0 x_t.
\end{align}
We refer the interested reader to Appendix~\ref{ax:ptdf_derivation} for a derivation of the PTDF matrix.
Superscript $0$ denotes the base-case PTDF, i.e. the pre-contingency case with no unplanned outages.

\subsection{Contingency Preliminaries}
\label{ssec:contingency_preliminaries}
Consider a contingency scenario $c$ such that $\set{L}_c \subseteq \set{L}$ is the set of one or multiple lines that experience an unplanned outage.
The post-contingency flow along any line $l \notin \set{L}_c$ is determined by line outage distribution factor $\LODF_{l\set{L}_c} \in \mathbb{R}^{1\times|\set{L}_c|}$ such that:
\begin{align}
    f^{c}_{t,l} = f^0_{t,l} + \LODF_{l\set{L}_c} f^0_{t,\set{L}_c},
    \label{eq:lodf_application}
\end{align}
where $f^{c}_{t,l}$ is the flow on line~$l$ in outage scenario~$c$, and $f^0_{t,l}$ is the pre-contingency flow on line~$l$ and $f^0_{t,\set{L}_c}$ is the vector of pre-contingency line flows of lines~$\set{L}_c$ at time~$t$. 
Note that any entry in $\LODF_{l\set{L}_c}$ can be either positive or negative. 
For the derivations of the LODFs we refer the interested reader to Appendix~\ref{ax:lodf_derivation}.
For every possible contingency $c \in \set{C}$ indexed by $c=\{1,...,C\}$ we use these sensitivity factors to define contingency-PTDF matrices~$B^c$ as:
\begin{align}
     B^c = B^0 + \begin{bmatrix}
        \LODF_{1\set{L}_c}  B^0_{\set{L}_c} \\
        \LODF_{2\set{L}_c}  B^0_{\set{L}_c} \\
        \vdots \\
        \LODF_{L\set{L}_c}  B^0_{\set{L}_c}
     \end{bmatrix},
     \quad \forall c \in \set{C},
     \label{eq:contingency_ptdfs}
\end{align}
where $B^0_{\set{L}_c}$ is the $|\set{L}_c|\times N$ matrix collecting the rows of $B^0$ corresponding to the outages in $c$.
Given a vector of nodal injections $x_t$ the resulting post-contingency power flows after outage $c$ can be computed as:
\begin{align}
    f_t^c = B^c x_t.
\end{align}
Note that the formulations and methodologies in this paper can be extended to accommodate generator contingencies, if they allow a linear representation, e.g. as shown in \cite{madani2016constraint}.

\subsection{Security Constrained Optimal Power Flow}\label{ssec:prelim_scopf}

We consider a multi-period preventive OPF as:
\allowdisplaybreaks
\begin{subequations}
\begin{align}
&\min_{g, x} \sum_{t\in\set{T}}C(g_t) \label{eq:cost_objective} \\
\text{s.t.} \qquad    
    & d_t + Mg_t  = x_t && \forall t \in \set{T}  \label{eq:nodal_eb} \\ 
    & e^\top x_t  = 0 && \forall t \in \set{T} \label{eq:global_eb} \\
    & \underline{g}_t \leq g_t \leq \overline{g}_t && \forall t \in \set{T} \label{eq:gen_cap} \\
    & - \overline{f}^0 \leq B^0 x_t \leq \overline{f}^0 && \forall t \in \set{T} \label{eq:loadflow_eq} \\
    & - \overline{f}^c \leq B^c x_t \leq \overline{f}^c && \forall t \in \set{T}, \forall c \in \set{C}. \label{eq:loadflow_cont_eq}
\end{align}%
\label{mod:general_SCOPF}%
\end{subequations}%
\allowdisplaybreaks[0]%
\noindent
Objective \cref{eq:cost_objective} minimizes the cost of generation given by cost-function $C(g_t)$.
Eqs.~\cref{eq:nodal_eb,eq:global_eb} enforce the nodal and global power balances.
Eq.~\cref{eq:gen_cap} imposes limits on the active power output of the generators. 
Eqs.~\cref{eq:loadflow_eq,eq:loadflow_cont_eq} enforce that no line is overloaded due to the resulting power flow for the base case and every contingency.
Thus, \cref{eq:loadflow_eq,eq:loadflow_cont_eq} define the feasible region of nodal injections given base and contingency PTDFs, and the thermal line flow limits:
\begin{align}
    \set{F}(B,\overline{f}) = \{x: -\overline{f} \leq B x \leq \overline{f}\},
    \label{eq:general_F}
\end{align}
where 
\begin{align}
    B = \begin{bmatrix}B^0 \\ B^1 \\ \vdots \\ B^C \end{bmatrix}, \quad
    \bar{f} = \begin{bmatrix}\bar{f}^0 \\ \bar{f}^1 \\ \vdots \\ \bar{f}^C \end{bmatrix}. 
    \label{eq:BfinF}
\end{align}
Using \cref{eq:general_F}, \cref{eq:BfinF} the following formulation is equivalent to \eqref{mod:general_SCOPF}:
\begin{subequations}
\begin{align}
&\min_{} \sum_{t\in\set{T}}C(g_t) \\
\text{s.t.} \qquad    
    & \text{\cref{eq:nodal_eb,eq:global_eb,eq:gen_cap}} \label{eq:all_other_constraints}\\
    & x_t \in \set{F}(B,\overline{f}) && \forall t \in \set{T}. \label{eq:flow_region_constraint}
\end{align}%
\label{mod:more_general_SCOPF}%
\end{subequations}%
\begin{remark}
\label{rem:possibleextension}
    As \cref{eq:flow_region_constraint} constrains the vector of nodal injections $x_t$ to be within the feasible region defined by the network capacity and its contingency scenarios, it is independent from both the objective function and other constraints on $x_t$. 
    Here the constraints in \cref{eq:all_other_constraints} only capture the power balances and the technical generator constraints. 
    However, other constraints e.g. inter-temporal constraints (e.g. to model storages) or generation related binary variables (e.g. to model unit-commitment) can be included without affecting \cref{eq:flow_region_constraint}.
\end{remark}

To reflect the upper and lower bounds of feasible region $\set{F}(B,\overline{f})$, each PTDF matrix $B^{0},\ldots,B^{C}$ introduces $2L$ linear inequalities to the problem.
Thus, even the least complex set of \mbox{N-1} contingencies, i.e. only one simultaneous outage, requires $2L(L+1)$ inequalities to define feasible region $\set{F}(B,\overline{f})$.
Furthermore, this set of inequalities has to be evaluated for every time step $t$ to solve \cref{mod:general_SCOPF}.
Therefore, the resulting problem size quickly becomes computationally intractable with increasing system size and more complex contingency scenarios. 
However, it has been shown that only a subset of these inequalities is necessary to sufficiently define $\set{F}(B,\overline{f})$, \cite{bouffard2005umbrella}, thus reducing computational complexity. 
In the following section we propose a procedure that discovers the minimal set of inequalities (constraints) based on endogenous model parameters and exogenous data characteristics.

\section{Redundancy Screening}
\label{sec:redundancy_screening}

\begin{figure}
    \centering
    \includegraphics[width=0.95\linewidth]{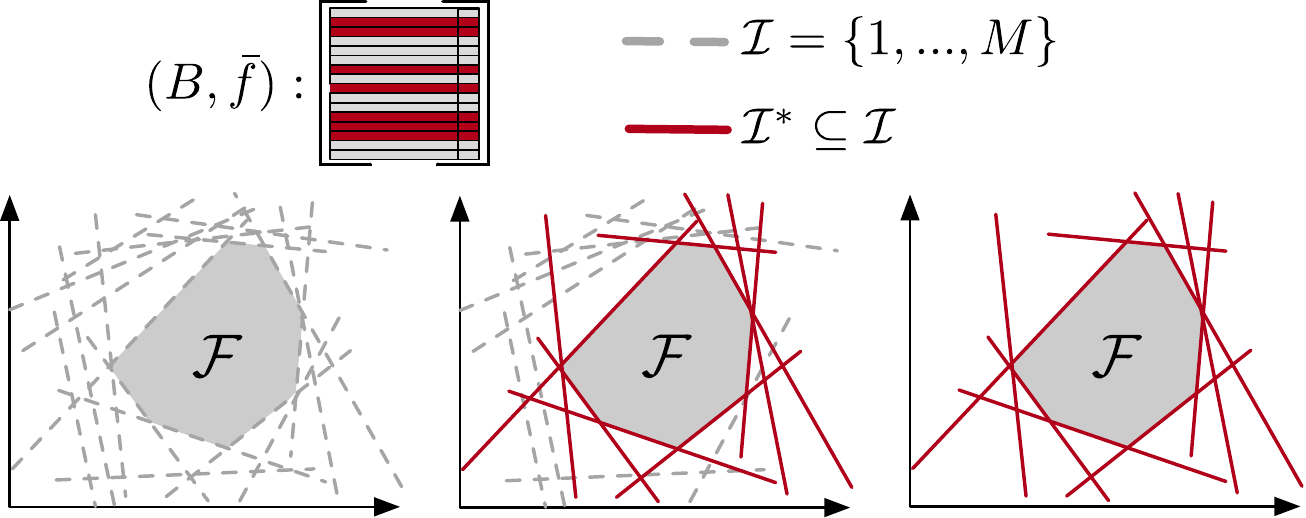}
    \caption{Schematic representation of the equivalent description of a feasible region
    $\set{F}(B,\overline{f},\set{I})$ by $\set{F}(B,\overline{f},\set{I}^*)$ where $\set{I}=\{1,...,M\}$ is the set of all indices of system $(B,\bar{f})$ and $\set{I}^*\subseteq\set{I}$ is the essential set of indices.}
    \label{fig:redundancy_pur_les_nuls}
\end{figure}

\begin{figure*}[t]
    \begin{minipage}{0.5\textwidth}
        \centering
        \includegraphics[width=0.95\linewidth]{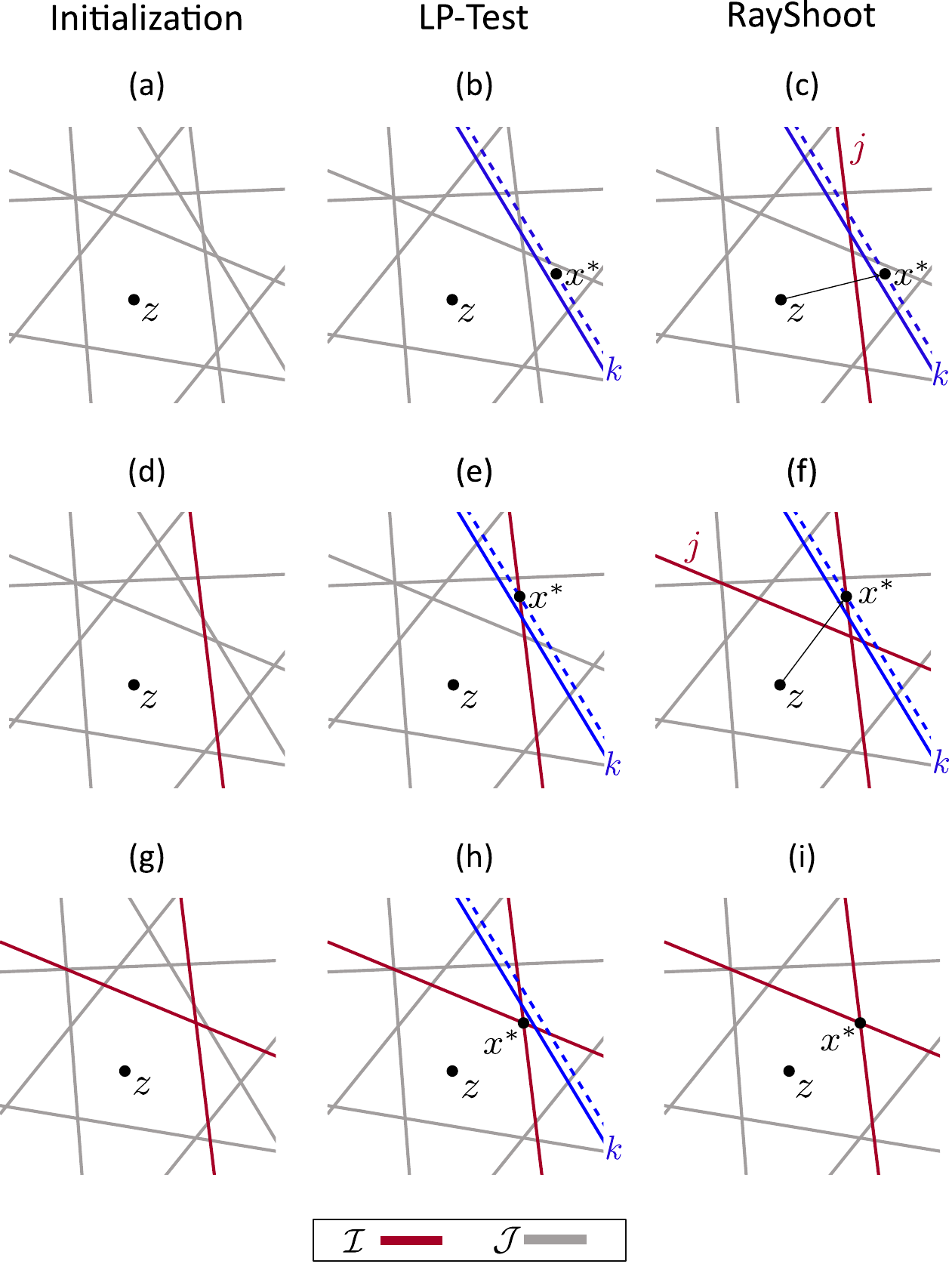}
    \end{minipage}%
    \begin{minipage}{0.5\textwidth}
    \removelatexerror % lol
    \begin{algorithm}[H]
        \label{alg:full_full_algorithm}
        \SetAlgoLined
        \SetKwFunction{RayShoot}{RayShoot}
        \SetKwFunction{Clarkson}{Clarkson}
        \SetKwFunction{LPTest}{LPTest}
        \SetKwInOut{Input}{input}\SetKwInOut{Output}{output}
        \Input{System of inequalities given by $B$ and $\overline{f}$, \\
               Interior point $z$}
        \Output{Returns the set $\set{I}$ of non-redundant inequality indices of the system $B x\leq \overline{f}$}
        \Begin{
          $\set{I} \leftarrow \emptyset$; \tcp{Set of essential indices}
          $\set{J} \leftarrow \{1,...,M\}$; \tcp{Indices to check}
          \While{$|\set{J}|>0$}{
            select an index $k$ from $\set{J}$\;
             $(p^*, x^*) \leftarrow  \text{\textbf{solve} LPTest}(B,\overline{f},\set{I} \cup \{k\},k)$\;  
            \eIf{$\exists\mkern2mu x^*$ \textbf{and} $p^* > \overline{f}_k$}{
                $\alpha \leftarrow true$\;
                $j \leftarrow \RayShoot(B,\overline{f},z,x^*)$\;
                \tcp*[h]{Returns an essential index}
            }{
                $\alpha \leftarrow false$;
            }
            \eIf{$\alpha$}{
                $\set{I} \leftarrow \set{I} \cup \{j\}$; \tcp*[h]{Update essential} \\
                $\set{J} \leftarrow \set{J} \setminus \{j\}$; \tcp*[h]{Remove checked}
            }{
                $\set{J} \leftarrow \set{J} \setminus \{k\}$; \tcp*[h]{Remove checked }
            }
            }
          \KwRet{$\set{I}$}
         }
        \caption{$\texttt{RedundancyRemoval}(B,\overline{f},z)$}
        \end{algorithm}
    \end{minipage}
    \caption{Graphical (left) and algorithmic (right) itemization of essential set discovery procedure, where each row in the graphic corresponds to one iteration step of the algorithm and each column corresponds to one specific task that is performed in each iteration (as given by the column headers). Gray lines represent the constraints with indices to be checked, red lines represent the found essential constraints, blue solid lines represent the constraint that is checked in the current iteration and blue dashed lines represent the corresponding relaxed constraint ($\overline{f}_k + 1$), see \eqref{eq:lp_test}.   
    (a) Initial state with $\set{J} = \{1,...,M\}$, $\set{I} = \emptyset$ and $z$ some interior point; (b) Some index $k$ is selected from $\set{J}$ and LP-Test($B,\overline{f},\set{I}, k$) is performed; 
    (c) Because $\set{I}$ is empty in the initial iteration $k$ is always non-redundant against $\set{I}$ and the most restricting constraint $j$ in the direction of $(x^*-z)$ is added to $\set{I}$;
    (d) The next iteration starts with $\set{I}$ now containing one essential index;
    (e) Because $k$ was non-redundant in the last step, it remains selected and LP-Test($B,\overline{f},\set{I}, k$) is performed;  
    (f) Now, $k$ is again non-redundant against $\set{I}$ and the most restricting constraint $j$ in the direction of $(x^*-z)$ is added to $\set{I}$; 
    (g) The next iteration starts with $\set{I}$ now containing two essential indices;
    (h) Because $k$ was non-redundant in the last step, it remains selected and LP-Test($B,\overline{f},\set{I}, k$) is performed;
    (i) Index $k$ is now redundant against set $\set{I}$ and is therefore removed from set $\set{J}$;
    The procedure repeats until all elements have been removed from $\set{J}$.}
    \label{fig:algoillus}
\end{figure*}

We consider the set of feasible solutions (feasible region) $\set{F}(B,\overline{f})$ to a linear program (LP) defined by system $(B\in \mathbb{R}^{M\times N},\overline{f}\in\mathbb{R}^{M})$ with $M>N$ and set of indices $\set{I}$ such that:
\begin{align}
    \set{F}(B,\overline{f},\set{I}) = \{x\in\mathbb{R}^N: B_ix \leq \overline{f}_i, \forall i \in \set{I}\},
    \label{eq:feasible_region}
\end{align}
where $B_i$ is the $i$-th row of matrix $B$ and $\overline{f}_i$ is the $i$-th entry of vector $\overline{f}$.
If follows from \eqref{eq:feasible_region} that if $\set{I} = \{1,...,M\}$, then $\set{F}(B,\overline{f},\set{I}) = \{x: Bx \leq \overline{f}\}=\set{F}(B,\overline{f})$.

\begin{definition}[Non-redundant/Redundant Index]
\label{def:essential_index}
Index ${k \in \set{I}}$ is called \textit{non-redundant} against set of indices $\set{I}$ if  $\set{F}(B,\overline{f},\set{I})$ changes when index $k$ is removed from $\set{I}$:
\begin{align*}
    k \in \set{I} \text{ is non-redundant iff } \set{F}(B,\overline{f},\set{I} \setminus \{k\}) \neq \set{F}(B,\overline{f},\set{I}).
\end{align*}
In analogy, index $k \in \set{I}$ is called \textit{redundant} if $\set{F}(B,\overline{f},\set{I})$ does not change when $k$ is removed from $\set{I}$:
\begin{align*}
    k \in \set{I} \text{ is redundant iff }\set{F}(B,\overline{f},\set{I} \setminus \{k\}) = \set{F}(B,\overline{f},\set{I}).
\end{align*}
\end{definition}

\begin{definition}[Essential Set/Index]
\label{def:nonredundant_set}
A set of indices ${\set{I}^* \subseteq \{1,...,M\}}$ is called \textit{essential} to the system $(B,\overline{f})$ if it contains all indices that are non-redundant against all indices $\{1,...,M\}$ of this system. In other words, no $k \in \set{I}^*$ can be removed from $\set{I}^*$ without changing $\set{F}(B,\overline{f},\set{I}^*)$ and ${\set{F}(B,\overline{f},\set{I}^*) = \{x: Bx \leq \overline{f}\}}$.
Accordingly, any index $k\in\set{I}^*$ is called essential index.
\end{definition}
\noindent
Fig.~\ref{fig:redundancy_pur_les_nuls} schematically illustrates a region $\set{F}(B,\overline{f})$ defined by a redundant system $(B,\bar{f})$ and indicates the relation between essential and non-essential indices.

\subsection{Essential Set Identification}
\label{ssec:essential_set_identification}

To identify essential set $\set{I}^*$, first we require a procedure that determines whether or not index $k$ is redundant in $\set{F}(B,\overline{f},\set{I})$.
Following \cite[Proposition 8.5]{fukuda2016lecture}, $k \in \set{I}$ is non-redundant if and only if the optimal solution $x^*$ and the corresponding optimal value $p^*$ of the LP:
\begin{subequations}\label{eq:redundancy}
\begin{align}
     \text{LP-Test($B,\overline{f},\set{I}, k$):} \quad & p^* = \max_x B_k x \\
     \text{s.t.} \quad   & B_i x \leq \overline{f}_i \quad \forall i \in \set{I} \setminus \{k\} \\ 
      & B_k x \leq \overline{f}_k + 1
\end{align}%
\label{eq:lp_test}%
\end{subequations}%
is strictly greater than $\overline{f}_k$.
Note that LP-Test will always find an optimal solution since set $\set{F}\neq\emptyset$ because it always contains at least $0$.
Using the LP-Test as given in \cref{eq:lp_test}, it is possible to identify essential set $\set{I}^*$ by running {LP-Test($B,\overline{f},\set{I}, k$)} with $\set{I} = \{1,...,M\}$ for all $k \in \set{I}$.
However, this requires solving a $M$-dimensional LP $M$ times.
This complexity can be significantly reduced by populating $\set{I}$ iteratively with identified essential indices, instead of always checking against complete set $\set{I} = \{1,...,M\}$, \cite{Szedlak_redundancy_removal_thesis,clarkson1994more}. 

The resulting iterative process \texttt{RedundancyRemoval} is illustrated in Fig.~\ref{fig:algoillus}.
The procedure takes system $(B,\bar{f})$ and an interior point $z\in \set{F}^\circ(B,\overline{f})$ as input and returns set $\set{I}^*$ of essential indices of system $(B,\bar{f})$.
Here $\set{F}(B,\overline{f})$ defines the feasible region of nodal injection vectors with respect to transmission limits and contingency scenarios. 
Therefore, $z=0$ will always be a point in the interior of this region, because zero nodal injections and thus zero-flows are always a solution to the power flow equations. 
The procedure is initialized with empty set $\set{I}=\emptyset$, which is iteratively filled with essential indices, and the full set $\set{J}=\{1,...,M\}$, which stores all indices that have to be checked. 
First, the procedure randomly selects an unchecked index $k$ from $\set{J}$ and solves the \mbox{LP-Test$(B,\overline{f},\set{I} \cup \{k\},k)$}, which returns $p^*$ and $x^*$ as per \cref{eq:lp_test}.
If the LP-Test returns an objective value $p^* > \bar{f}_k$, then $\set{I}$ does not yet contain the index of a constraint that restricts $\set{F}(B,\bar{f},\set{I})$ in the direction of $x^*-z$, see Fig.~\ref{fig:algoillus}b).
However, because set $\set{I}$ is initialized empty, indices can be non-redundant against $\set{I}$ but not essential to $(B,\overline{f})$. 
In other words, there might exist a constraint with index $j$ in the direction of $x^*-z$ that is more restrictive than the constraint with index $k$. 
As shown in Fig.~\ref{fig:algoillus}c), the auxiliary procedure \texttt{RayShoot} identifies this most restrictive constraint in the direction of $x^*-z$ by shooting a ray from $z$ in the direction of $x^*-z$ and returning index $j$ of the first hyperplane $\{x : B_j x = \overline{f}_j\}$ that it crosses. 
This index $j$ is guaranteed to be an essential index of $(B,\bar{f})$ and is thus added to $\set{I}$ and removed from $\set{J}$.
See Appendix~\ref{ax:rayshoot} for a detailed description of \texttt{RayShoot}.
Note that if $j\neq k$, then $k$ remains in $\set{J}$ to be checked again, see Fig.~\ref{fig:algoillus}e).
If \mbox{LP-Test$(B,\overline{f},\set{I} \cup \{k\},k)$} determines $k$ to be redundant against $\set{I}$, see Fig.~\ref{fig:algoillus}h), then $k$ is guaranteed to be not essential because $\set{I}$ only contains essential indices. 
In this case, no new essential index has been found and $k$ is removed from $\set{J}$, see Fig.~\ref{fig:algoillus}i).
The process is repeated until $\set{J}$ is empty, thus guaranteeing a termination of the algorithm in finite time.
The resulting set $\set{I}$ contains all essential indices and therefore $\set{I}=\set{I}^*$, \cite[Theorem 2.2.1]{Szedlak_redundancy_removal_thesis}.
This essential set $\set{I}^*$ is a minimal representation of the contingency feasible region, see \eqref{eq:general_F}, and each essential index represents a specific critical line under a specific outage and therefore can be denoted as a minimal set of CBCOs.

While the complexity of \texttt{RedundancyRemoval} remains dominated by the LP-Test, it is now performed $M$ times with \textit{at most} $|\set{I}^*|$ constraints.
The worst-case performance of \texttt{RedundancyRemoval} occurs when all essential indices are found in the first $|\set{I}^*|$ iterations. 
Then, LP-Test is performed $|\set{I}^*|$ times with less than $|\set{I}^*|$ constraints and $M-|\set{I}^*|$ times with $|\set{I}^*|$ constraints. 
The \texttt{RayShoot} procedure performs basic vector calculations in the $\mathbb{R}^{M\times N}$ space and is performed $|\set{I}^*|$ times.
Thus, its complexity is linear against $MN$ and dominated by the complexity of \mbox{LP-Test}. 

\begin{remark}
\label{rem:symmetrical_calculations}
The capacity of a line is independent from the direction of the flow, which leads to 
identical constraints on the nodal injections for the upper and the lower bound but with reversed sign. 
Therefore, an essential set related to the upper bounds directly corresponds to an essential set for the lower bounds and it is sufficient to perform the \texttt{RedundancyRemoval} only on the positive PTDF matrices to speed-up the essential set identification.
\end{remark}

\subsection{Conditional Redundancy}
\label{ssec:conditional_redundancy}

\begin{figure}[t]
    \centering
    \includegraphics[width=0.7\linewidth]{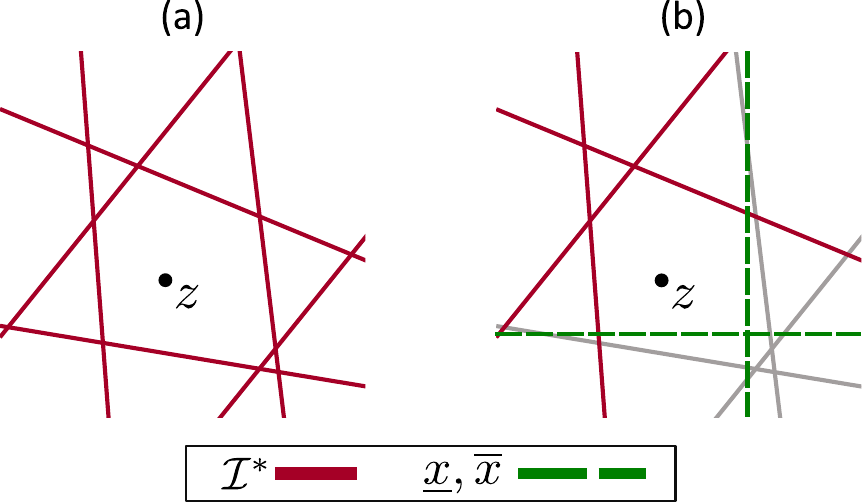}
    \caption{Illustration of two essential sets $\set{I^*}$ with and without considerations for bounds on $x$: a) $\set{F}(B,\overline{f},\set{I}^*) = \{x: Bx \leq \overline{f}\}$ b) $\set{F}(B,\overline{f},\set{I}^*|_{(\underline{x},\overline{x})}) = \{\underline{x} \leq x \leq \overline{x}: Bx \leq \overline{f}\}$ }
    \label{fig:cond_redundancy}
\end{figure}

The essential set identification as presented in previous Section~\ref{ssec:essential_set_identification} only depends on redundancies that are inherent to system $(B,\bar{f})$, i.e. that are given by the power flow limits and contingency scenarios as in \eqref{eq:general_F}--\eqref{mod:more_general_SCOPF}.
Thus, resulting essential set $\set{I}^*$ contains all non-redundant indices assuming that $x$ is unbounded. 
While it is useful to find such a general essential set, practical application usually includes specific generation units, demand- and renewable time-series along with the grid infrastructure. 
This allows to determine upper and lower bounds for nodal injections $x_t$. 
Considering bounds on $x_t$ in the proposed algorithm, can render certain essential indices unnecessary, because the specific allocation of nodal injections to overload certain CBCOs will never occur given the known technical limits. 
In other words, we can find a set $\set{I}^*|_{(\underline{x},\overline{x})}\subseteq \set{I}^*$ by bounding $x_t$ as schematically illustrated in Fig.~\ref{fig:cond_redundancy}.
Resulting set $\set{I}^*|_{(\underline{x},\overline{x})}$ is then sufficient to define $\set{F}(B,\overline{f})$ \textit{under the condition} that $x$ is bounded by $(\underline{x},\overline{x})$:
\begin{equation}
\begin{split}
 \set{F}(B,\overline{f},\set{I^*}) &= \set{F}(B,\overline{f},\set{I^*}|_{(\underline{x},\overline{x})}) \\
 &= \{\underline{x} \leq x \leq \overline{x}: Bx \leq \overline{f}\}.   
\end{split}
\end{equation}

Bounds $(\underline{x},\overline{x})$ strictly relate to the parameters and available data of the modeled system. 
In typical applications, the modeled system remains static over $\set{T}$, so that implicit bounds on nodal injections will always hold and a smaller essential set will provide a reduction of model complexity without compromising the validity of the resulting SCOPF.
First, we compute asymmetrical bounds by determining the maximum positive and negative nodal injections:
\begin{align}
    \check{x}_n &= \min(d_{t,n},~t\in\set{T}) + \min(M_n \underline{g}_t, t\in\set{T}) 
    \label{eq:assymetric_low_x} \\
    \hat{x}_n &= \max(M_n \overline{g}_t,~t\in\set{T}) 
    , \label{eq:assymetric_high_x}
\end{align}
where $\check{x}_n$ and $\hat{x}_n$ are the maximum negative and maximum positive nodal injection at $n$ given the available demand and generation parameters.
Note that these bounds can be extended to accommodate renewable in-feed time series or storage capacities. 
However, as indicated in Remark~\ref{rem:symmetrical_calculations}, feasible region $\set{F}(B,\overline{f})$ is symmetric. 
Thus, bounds on $x$ have to be included symmetrically and we define:
\begin{align}
    -\underline{x}_n = \overline{x}_n = \max(|\check{x_n}|, |\hat{x_n}|).
    \label{eq:symmetric_bounds}
\end{align}
Note that symmetric definition of the bound in \cref{eq:symmetric_bounds} also assures that $z=0$ remains an interior point of feasible region  $\set{F}(B,\overline{f},\set{I^*}|_{(\underline{x},\overline{x})})$.
Using these bound to compute $\set{I}^*|_{(\underline{x},\overline{x})}$ will further reduce the resulting problem size of the SCOPF \eqref{mod:more_general_SCOPF}.
Note that the identification of the conditional essential set uses the extreme (upper and lower) technical limits of all resources connected to a node. 
Therefore $\set{I}^*|_{(\underline{x},\overline{x})}$ can also be applied to solve SCOPF problems that impose additional constraints on the behavior of these resources, see Remark~\ref{rem:possibleextension}.

\section{Impact Screening}
\label{sec:lodf_filter}

The run-time of the \texttt{RedundancyRemoval} is directly related to the initial number of constraints $M$ since each index $k \in \{1,...,M\}$ has to be checked. 
It is therefore desirable to reduce the number of constraints beforehand if possible. 
As described in Section~\ref{ssec:contingency_preliminaries}, contingencies are considered by computing how line flows are distributed across all other lines in the case of an outage. 
Each line is only significantly affected by an outage of its physical neighbors in close proximity, while a large number of contingencies in greater electrical distance have hardly any effect on its post-contingency power flow. 

Consider the outage of a line $o \in \set{L}$. As per \cref{eq:lodf_application}, $\LODF_{lo}$ determines how the pre-contingency power flow of line $o$ is distributed among all other lines $l\neq o, l \in \set{L}$.
Because the power flow on any line is bounded by $\overline{f}$, the impact any line outage can have on any other line is bounded by the respective $\LODF$ multiplied with the maximum flow on this line:
\begin{align}
     |f_{t,l}^o - f^0_{t,l}| = |\LODF_{lo}  {f}^0_{t,o}| \leq |\LODF_{lo} \overline{f}_o|.
\end{align}
By reserving a small capacity margin $\eta$ on each line, every outage that impacts this line by less than $\eta$ can be disregarded. 
In other words, all rows $\PTDF^o_l$ can be omitted if
\begin{equation}
 \frac{\LODF_{lo}  \overline{f}_o}{\overline{f}_l} < \eta,
\end{equation}
effectively reducing the length of the input matrix $B$ and therefore reducing the run-time of the \texttt{RedundancyRemoval}. 

Depending on the implementation, chosen threshold $\eta$ either reflects a safety margin by reducing the available line capacity $(1-\eta)\overline{f}_l$ or an allowable \textit{worst-case} short-term overload by virtually increasing the line capacity $(1+\eta)\overline{f}_l$. 
Both approaches are typically used in practice to accommodate parameter uncertainty, \cite{bienstock2016electrical}.
Note that if $\eta$ is defined as a safety margin, the results of the SCOPF may be altered because less line capacity is available.

\section{Case Study}
\label{sec:case_study}

This section investigates the mechanics of the proposed constraint reduction process on two specific data sets. 
First, we solve the N-1 DC SCOPF for the IEEE 118 bus system with $186$ lines and line capacity information taken from \cite{pglib}.
This system is a suitable example to illustrate proposed methodology and allows comparability to related methods due of its common application, e.g. in \cite{ardakani2013identification,ardakani2014acceleration}.
Second we use a larger 453-node data set of the German transmission system (DE case) to showcase the performance for common real-world multi-period applications.
The DE case comprises almost 2 million constraints related to its $995$ lines.
Table~\ref{tab:overview_cases} summarizes both cases.

We show four stages of constraint reduction that have been used to solve the N-1 SCOPF.
Stage ``Full'' considers all combinations of branches and outages in the positive halfspace, i.e. no explicit constraint reduction has been applied beyond ignoring the symmetry of the flow limits as discussed in Remark~\ref{rem:symmetrical_calculations}. 
The ``Pre'' (preprocessed) stage includes the impact screening as described in Section~\ref{sec:lodf_filter}. 
The stages ``RR'' and ``CRR'' apply the \texttt{RedundancyRemoval} algorithm on the impact-screened N-1 PTDF without and with conditional redundancies, see Section~\ref{ssec:conditional_redundancy}.
Note that those stages are presented here to itemize the effect of the different parts of the reduction algorithm.
For actual application of the proposed method there is only one stage to use, i.e. ``CRR''. 
All results are compared in terms of the resulting number of constraints and the corresponding time to solve the optimal power flow model \eqref{mod:general_SCOPF} using these constraints.

\begin{remark}\label{rem:scopf_implementation}
The implementation of the SCOPF problem in this case study is based on the PTDF formulation in \eqref{mod:general_SCOPF}, where the used $B^c$ contingency PTDFs have been reduced to represent the minimal set of constraints given by essential set $\set{I}^*$.
However, other implementations of the SCOPF are possible, e.g using voltage angles or decomposition techniques, since the CBCOs given by essential set $\set{I}^*$ exactly define the lines and contingencies that have to be included.
\end{remark}

The computations have been performed on a standard PC workstation with an Intel 8th generation i5 processor and 16GB memory.
The optimal power flow model and the reduction procedures have been implemented in the open source \textit{Power Market Tool} (POMATO, \cite{pomato}). 
The tool is written in Python for data pre- and postprocessing and uses the Julia/JuMP package, \cite{DunningHuchetteLubin2017}, in combination with the Gurobi solver, \cite{gurobi}, as its optimization kernel.
To allow direct comparison, dual simplex was used and the presented times are the times reported by the solver, including presolve. 

\begin{table}[t]
\centering
\caption{Overview: Case Studies}
\label{tab:overview_cases}
\begin{tabular}{lllll}\toprule
            & Nodes & Lines & Generators & N-1 Flow Constraints \\\midrule 
IEEE 118    & 118   & 186   & 116   &  66,216\\
DE          & 453   & 995   & 4226  &  1,934,280\\  
\bottomrule
\end{tabular}
\end{table}

\subsection{IEEE 118 Bus Case}
\label{sec:ieee_118_case_study}

Table~\ref{tbl:results_ieee} itemizes the number of constraints, the respective solve times and objective values for all constraint reduction stages in the IEEE 118 bus case.
The OPF has been solved for a single time step.
For the preprocessing phase, the impact screening margin set to $\eta = \unit[5]{\%}$ which reduces the set of constraints by $\unit[87]{\%}$ to $4,152$ and thus reducing the solve time by $\unit[81]{\%}$. 
The small objective value increase (\textcolor{black}{approximately} $\unit[3]{\%}$) results from the implicit line capacity reduction of the impact screening margin, see Section~\ref{sec:lodf_filter_discussion} below.

Running \texttt{RedundancyRemoval} further reduces this set by $41\%$ to $2,465$ and including conditional redundancy, as described in Section \ref{ssec:conditional_redundancy}, yields a set of only $518$ CBCOs, that guarantee a N-1 SCOPF.
Thus, instead of $L$ relevant contingencies for $L$ lines we observe an average of $2.78$ critical outages per line.
This represents a total removal of over $\unit[98]{\%}$ of the constraints and results in a $\unit[97]{\%}$ reduction of the time needed to solve the problem.
Furthermore, we observe that the objective value remains unchanged after the impact screening. This verifies, that the reduction due to Algorithm~\ref{alg:full_full_algorithm} indeed only removes redundant constraints.
The process time of \texttt{RedundancyRemoval} for the 118 bus case is around \unit[7]{min} without and below \unit[1]{min} with conditional redundancy. 
This demonstrates, that the process time of the presented algorithm 
reduces the more redundant the system is, i.e. the fewer non-redundant constraints can be found.
{\color{black} % Block process time
Note that the reported process times of the ``Full'' and ``Pre'' stages reflect the time needed to calculate the {N-1~PTDF} matrices only, while the process times of the ``RR'' and ``CRR'' stages also include the run time of the \texttt{RedundancyRemoval}.
The process time is lower in the ``Pre'' stage, since the N-1~PTDF computation is integrated with the impact screening methodology and, thus, a smaller PTDF matrix is generated. 
}
When using the full \mbox{N-1~PTDF} as the input to the \texttt{RedundancyRemoval}, we find a potentially larger set of CBCOs in significant more process time (RR: 3265 constraints in \unit[1216]{s}, CRR: 518 constraints in \unit[171]{s}), both guaranteeing SCOPF with the same objective value as the ``Full'' case as the full line capacity is available. This example shows that the technical limits imposed by the ``CRR'' stage are often more restrictive than the impact screening, leading to the same essential set with and without impact screening. 

The reduction in solution time does not match the reduction in constraints. Since the SCOPF for the IEEE 118 case study is solved, in contrast to common economic applications, for a single time step since the IEEE cases do not come with time series, the major advantage that each market clearing profits from the preprocessing, does not apply.

\begin{table}
\caption{IEEE 118 bus case constraint and solve time reduction}
\label{tbl:results_ieee}
\centering
\begin{tabular}{rllll}\toprule
                               & Full               & Pre         & RR          & CRR  \\ \midrule
\# Constraints                 & 33,108             & 4,152       & 2,465       & 518         \\
Process Time [s]               & 1.22               & 0.215       & 396         & 64.9       \\
Presolve [s]                   & 3.84               & 0.45        & 0.27        & 0.07        \\
Solve Time [s]                 & 7.13               & 1.34        & 0.76        & 0.23        \\
Objective                      & 119,996            & 124,103     & 124,103     & 124,103     \\ \midrule
\multicolumn{2}{r}{total constraint reduction\textsuperscript{*}:}      & 87\%        & 93\%        & 98\%        \\
\multicolumn{2}{r}{additional constraint reduction\textsuperscript{*}:} & 87\%        & 41\%        & 79\%        \\
\multicolumn{2}{r}{total solve time reduction\textsuperscript{*}:}       & 81\%        & 89\%        & 97\%        \\
\bottomrule\\[-6pt]
\multicolumn{2}{l}{\textsuperscript{*}Relative to ``Full''} &   &   & 
\end{tabular}
\end{table}

\subsection{DE Case}

The DE case solves single and multi-period nodal market clearing for the German power system including inter-temporal constraints for energy storages.
The power plant data is based on \cite{opsd_2018} and the spacial distribution and grid topology is based on \cite{kunz_electricity_2017}. 
The large set of power plants is due to a detailed regionalization of small scale, decentralized power plants.
This case represents a real world application with a prohibitively large linear problem. 
Indeed, the the full set of constraints cannot be solved by the computer hardware used for this case study as the system runs out of memory before an optimal solution has been obtained. 
While approaching the problem with more powerful hardware might be able to overcome this, the application of the proposed redundancy removal procedures makes this problem solvable. 
Table~\ref{tbl:results_de_constraints} shows that the ``CRR'' method removes $\unit[99.7]{\%}$ of all constraints within $\unit[195]{min}$ processing time.
The resulting average number of critical outages per line is $2.64$ which is surprisingly similar to the 118 bus case.
Preprocessing alone reduces the number of constraints already by over $\unit[98]{\%}$ with a impact screening margin of $\eta = \unit[5]{\%}$. 
An additional $\unit[26]{\%}$, $\unit[75]{\%}$ are achieved by the \texttt{RedundancyRemoval} without and with conditional redundancy, respectively.
Again, the processing time of the \texttt{RedundancyRemoval} itself is larger in the ``RR'' stage relative to ``CRR'' as the problem is less redundant. 
The resulting set of CBCOs is used to solve the SCOPF for a single time step and two time series of 10 and 24 time steps. The 24 time steps are the hours of an arbitrarily chosen day in January 2017, with the 10 time steps being the first ten. In the 24 time step run, 26 different line and contingency constraints are active. Since the bounds for the conditional redundancy have been determined for the whole year, the set of CBCOs will guarantee a contingency secure solution for all time steps, however with a potentially different set of active constraints.
The solve times and objective values are itemized in Tables~\ref{tab:de_time_10} and \ref{tab:de_time_24}.
As there is no data for the ``Full'' stage, the time reductions are reported relative to the ``Pre'' stage.
While the problem was not solvable with the full set of constraints, after ``CRR'' reduction an optimal solution was found within $\unit[3.2]{s}$ for the single time step run, $\unit[13.5]{s}$ for the 10-time step run and $\unit[22.87]{s}$ for the 24-time step run.

Compared to the single and 10-time step run, the 24-time step run shows a higher total time reduction both absolute as well as relative to the constraint reduction.
This highlights the positive effect of the larger time series, where the benefits of the constraint reduction apply in every time step. 
All stages in the two runs result in exactly the same objective value verifying the the removal of only redundant constraints.

\begin{table}
\caption{DE case constraint and solve time reduction (single time step)}
\label{tbl:results_de_constraints}
\centering
\begin{tabular}{rllll}\toprule
                    & Full                           & Pre                       & RR          & CRR         \\ 
\midrule
\# Constraints      & 967,140                        & 14,523                    & 10,695      & 2,629       \\
Process Time [s]    & 134                            & 2.62                      & 136,495     & 11,719      \\
Presolve [s]        & NA                             & 7.56                      & 5.73        & 0.77        \\
Solve Time [s]      & NA                             & 12.61                     & 10.76       & 3.20        \\
Objective           & NA                             & 726,505                   & 726,505     & 726,505     \\ \midrule
\multicolumn{2}{r}{total constraint reduction\textsuperscript{*}:}      & 98.5\%                    & 98.9\%      & 99.7\%      \\
\multicolumn{2}{r}{additional constraint reduction\textsuperscript{*}:} & 98.5\%                    & 26\%        & 75\%        \\
\multicolumn{2}{r}{total solve time reduction\textsuperscript{*}:}      &                           & 15\%        & 68\%        \\ 
\bottomrule\\[-6pt]
\multicolumn{2}{l}{\textsuperscript{*}Relative to ``Pre''} &   &   & 
\end{tabular}
\end{table}

\begin{table}[t]
\centering
\caption{DE case solve time reduction  (10 time steps)}
\label{tab:de_time_10}
\begin{tabular}{rllll}
\toprule
                    & Full          & Pre             & RR          & CRR   \\ 
\midrule
Presolve [s]        & NA            & 93.4            & 84.41       & 13.5   \\
Solve Time [s]       & NA            & 283.19          & 250.29      & 44.82   \\
Objective           & NA            & 2,993,021       & 2,993,021   & 2,993,021    \\
\midrule
\multicolumn{3}{r}{total solve time reduction\textsuperscript{*}:}      & 12\%       & 84\%       \\
\bottomrule\\[-6pt]
\multicolumn{2}{l}{\textsuperscript{*}Relative to ``Pre''} &   &   & 
\end{tabular}
\end{table}

\begin{table}[t]
\centering
\caption{DE case solve time reduction (24 time steps)}
\label{tab:de_time_24}
\begin{tabular}{rllll}
\toprule
                  & Full        & Pre               & RR          & CRR    \\ 
\midrule
Presolve [s]      & NA          & 507.7             & 220.53      & 22.87   \\
Solve Time [s]     & NA          & 1,707.56          & 714.37     & 89.53   \\
Objective         & NA          & 8,764,696         & 8,764,696   & 8,764,696    \\
\midrule
\multicolumn{3}{r}{total solve time reduction\textsuperscript{*}:}      & 58\%        & 95\%        \\
\bottomrule\\[-6pt]
\multicolumn{2}{l}{\textsuperscript{*}Relative to ``Pre''} &   &   & 
\end{tabular}
\end{table}

\subsection{Impact Screening}
\label{sec:lodf_filter_discussion}

\begin{figure}
    \centering
    \definecolor{mycol}{HTML}{250ADC}
\definecolor{mycol_red}{HTML}{A50026}

\begin{tikzpicture} 
\begin{axis}[
        width=\linewidth,
        height=0.3\textwidth,
        axis y line*=left,
        grid=major, 
        xmin=0, xmax=11,
        ymin=0, ymax=60000,
        scaled y ticks = base 10:-3,
        xlabel= $\eta$ in \%, 
        xtick={0,1,2,3,4,5,6,7,8,9,10},
        ytick={0,10000,20000,30000,40000,50000,60000},
        ylabel=Number of Constraints,
        label style={font=\small},
        tick label style={font=\small} 
        ]
        \addplot[color=mycol, 
            line width=1pt,
            mark size=2pt,
            mark=*] 
            coordinates { 
            (1.0, 53720)
            (1.5, 39806)
            (2.0, 31729)
            (2.5, 26306)
            (3.0, 22602)
            (3.5, 19799)
            (4.0, 17608)
            (4.5, 15997)
            (5.0, 14523)
            (5.5, 13312)
            (6.0, 12279)
            (6.5, 11499)
            (7.0, 10773)
            (7.5, 10121)
            (8.0, 9539)
            (8.5, 8997)
            (9.0, 8550)
            (9.5, 8148)
            (10.0, 7782)
            % (10.5, 7466)
            }; \label{plot_one}
            
    % \addlegendentry{{CBCOs}}    
    \end{axis}
        
    \begin{axis}[
                width=\linewidth,
                height=0.3\textwidth,
                axis y line*=right,
                scaled ticks = true,
                ylabel near ticks,
                ymin=2.7e6, ymax=3.3e6,
                xmin=0, xmax=11,
                x axis line style={draw=none},
                xtick style={draw=none},
                xtick=\empty,
                label style={font=\small},
                tick label style={font=\small}, 
                legend style={font=\small},
                ylabel=Objective Value,
                ylabel shift=-10pt,
                legend columns=2,] 
        \addlegendimage{/pgfplots/refstyle=plot_one}\addlegendentry{\#Constraints}
        \addplot[color=mycol_red, 
            line width=1pt,
            mark size=2pt,
            mark=square*]  
            coordinates { 
            (1.0, 2931725)
            (2.0, 2946841)
            (3.0, 2962108)
            (4.0, 2977517)
            (5.0, 2993021)
            (6.0, 3008684)
            (7.0, 3024489)
            (8.0, 3040418)
            (9.0, 3056472)
            (10.0, 3072641)
            }; 
    \addlegendentry{Objective}
    \end{axis}
\end{tikzpicture}
\caption{Effect of impact screening margin $\eta$ on the resulting number of constraints for the DE 10-time step case.}
\label{fig:cb_to_sensitivity_plot}
\end{figure}
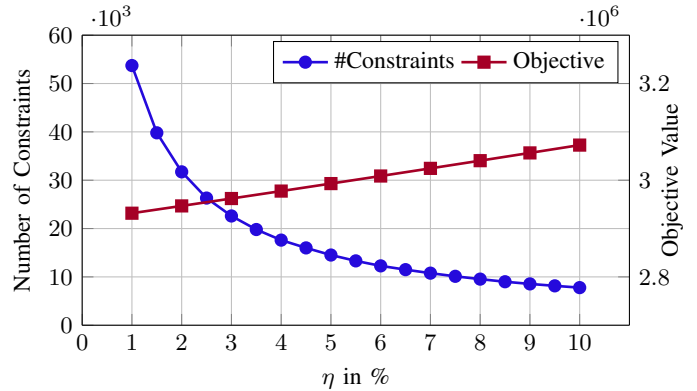

As described in Section~\ref{sec:lodf_filter}, the impact screening implicitly reduces the available line capacity in favor of disregarding outages which can not exceed this margin in case of an outage.
While this significantly reduces the number of considered contingencies, the available transfer capacity of the network is reduced. 
The reduced network capacity correlates with a higher objective value as cheaper generators are more restricted to supply electrically distant nodes.
To itemize the effect of the choice of the margin $\eta$, the DE case 10-time step run was repeated with different settings for $\eta$, where no further reduction was applied. 
Fig.~\ref{fig:cb_to_sensitivity_plot} shows the number of constraints resulting from the impact screening, and objective values in the optimal solution.
Note that the $\eta=\unit[5]{\%}$ data-point in Fig.~\ref{fig:cb_to_sensitivity_plot} matches the ``Pre'' stage from Table~\ref{tab:de_time_10}.

\begin{figure}
    \centering
    \includegraphics[width=0.95\linewidth]{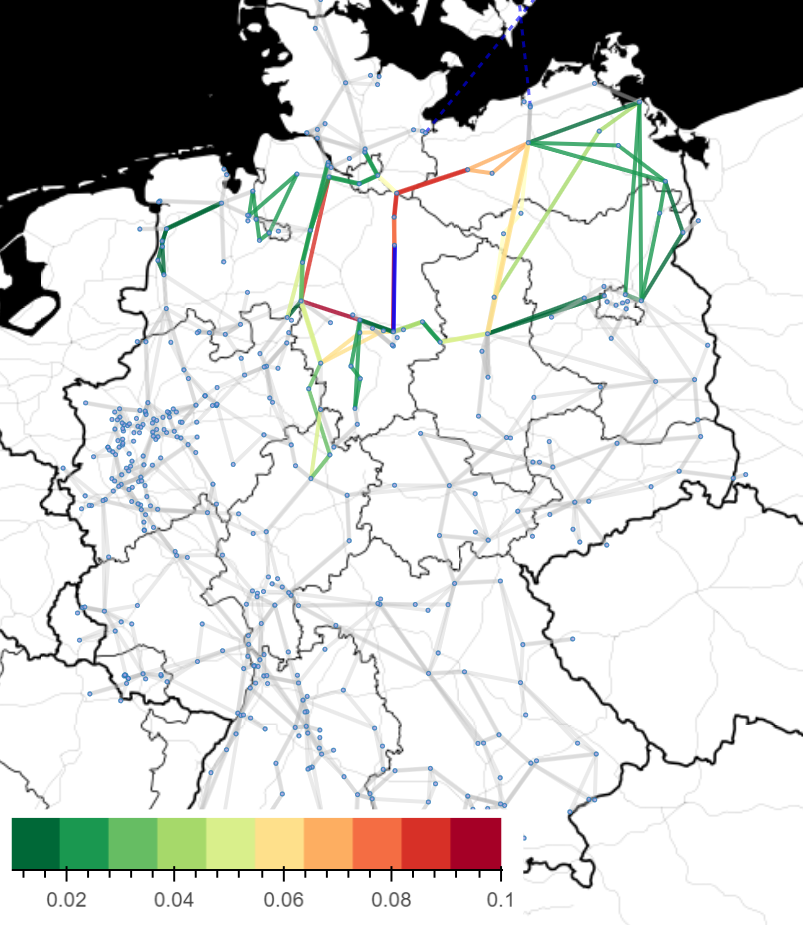}
    \caption{Impact of outages towards the highlighted (blue) line; Grey lines indicate a sensitivity of less than \unit[1]{\%}}.
    \label{fig:lodf_de}
\end{figure}

The effect of $\eta$ on the objective is closely linear and we observe that an increase in $\eta$ of \unit[1]{\%} translates into a mild increase of the objective value of \textcolor{black}{approximately} \unit[0.5]{\%}.
On the other hand, the resulting number of constraints is reduced drastically already by small values of $\eta$.
Those results highlight how every outage in a meshed grid only has a certain reach and that the number of outages relevant for a specific branch is spatially restricted.
Fig.~\ref{fig:lodf_de} shows this effect by color-coding the relative outage sensitivity of all lines in the network towards the highlighted blue line. 
By showing all lines with an impact of less than \unit[1]{\%} in gray, we see that mostly neighboring and parallel lines in close proximity have a significant impact on the highlighted line.

\section{Larger Test Cases and Scalability}

This section investigates the scalability of the method and discusses some considerations for practical implementation. 
Here we used the 1159 bus European (CWE) data-set from \cite{schonheit2020impact} and the A\&M synthetic 2000 bus network (ACTIVSg2000) from \cite{birchfield_grid_2017}.
Our experiments showed that a direct implementation of the described algorithm is able to find a solution also for larger test cases, but exceeds desirable time frames.
However, using a more effective execution of the proposed algorithm, process times are reduced to a reasonable level without changing the proposed method itself. 
Notably, the inherent sequential nature of \texttt{RedundancyRemoval} obstructs parallel computation, because every constraint is checked against both already identified essential indices and unchecked indices, see Fig.~\ref{fig:algoillus}.
However, it is possible to run the algorithm multiple times in parallel using segments of the full set of indices. 
Indices found redundant in a segment are also redundant against the whole set and can thus be removed. 
However, indices identified as non-redundant in a segment are not necessarily essential with respect to the whole set and therefore need to be either confirmed or discarded as an essential index by running a final instance of the algorithm with all remaining indices.
Note that the segmentation does not alter the algorithm as described in Section~\ref{sec:redundancy_screening}, nor affects the final result of the \texttt{RedundancyRemoval}, but only reduces the process time.

For both of the larger cases, with parameters as reported in Table~\ref{tab:large_cases}, we applied the ``CRR'' method with impact screening as described in Section~\ref{sec:case_study}.
The resulting number of CBCOs, the process time of the algorithm and the solve time of the SCOPF (in the implementation described in Remark~\ref{rem:scopf_implementation}) are itemized in Table~\ref{tab:large_cases_results}.
In both cases, the resulting CBCOs constitute a constraint reduction of over \unit[99.9]{\%} and enabled solving an SCOPF of the CWE case in \unit[13.96]{s} and the A\&M 2000 bus case in \unit[27]{min}.
Notably, as the the 2000 bus network contains a larger number of medium voltage lines with lower thermal rating, fewer constraints can be considered redundant.
This leads to an over proportional increase in process and solve time, considering that the 2000 bus network has only \unit[30]{\%} more lines than the CWE case.
{\color{black} % Block More effevtive
Further, comparing the process times of the DE case (\unit[11,719]{s}, see Table~\ref{tbl:results_de_constraints}) with the larger CWE case (\unit[3,906]{s}, see Table~\ref{tab:large_cases_results}) illustrates how significant performance improvements can be achieved with more effective execution of the algorithm as outlined at the beginning of this section.
}

\begin{table}[t]
\centering
\caption{Overview: Larger Test Cases}
\label{tab:large_cases}
{\color{black}
\begin{tabular}{lllll}\toprule
                & Nodes  & Lines     & Generators   & N-1 Flow Constraints \\\midrule 
CWE             & 1,159  & 2,438     &  4,797       &  11,278,188    \\
A\&M 2000       & 2,000  & 3,206     &  3,535       &  17,671,472    \\  
\bottomrule
\end{tabular}
}
\end{table}

\begin{table}[t]
\centering
\caption{Larger Test Cases Results (using ``CRR'')}
\label{tab:large_cases_results}
{\color{black}
\begin{tabular}{llll}\toprule
                & \# Constraints     & Process Time [s]  & Solve Time [s]  \\\midrule 
CWE             & 5,349             & 3,906             & 13.96           \\
A\&M 2000       & 10,869            & 66,735            & 1,676           \\  
\bottomrule
\end{tabular}
}
\end{table}

\section{Conclusion}

This paper proposed a methodology to identify the minimal set of constraints that define the space of feasible nodal injections in an electricity transmission network with contingency scenarios. 
This set of critical branches under critical outages (CBCOs) can be used to significantly reduce the dimensionality, and thus computational complexity, of security-constrained optimal power flow (SCOPF) problems. 

First, we presented an algorithm that identifies the indices of the constraints that define the CBCOs, for a given system of linear inequalities.
This procedure yields feasible process times by iteratively selecting only constraints that correspond to the innermost hyperplanes defining the solution space and, therefore, are non-redundant.
Second, we proposed two methods to further reduce the run time of the algorithm and the resulting number of CBCOs by internalizing technical limits of the nodal injections and enforcing a minimal worst-case impact of contingencies to line flows. 

The proposed algorithm has been applied to solve SCOPF problems for the IEEE 118 bus system, the A\&M synthetic 2000 bus system, as well as two real-world data sets of the German and European transmission system. 
The algorithm is shown to return all CBCOs within reasonable time (within minutes for smaller cases and hours for the larger cases) and for every case at least \unit[95]{\%} of the constraints are identified as redundant.
Using the identified set of CBCOs to solve SCOPF problems for these networks showed significant improvements in solve time. 
For example, a single DC SCOPF for the IEEE 118 bus system has been solved in \unit[0.23]{s} and the German data set has been solved for 24 time steps in less then \unit[90]{s}. 
All code and input data have been published supplementary to this paper as open-source software. 

Considerations for practical implementation and solving larger cases have been discussed. 
For the presented data-sets the resulting solve times are feasible for the intended application of enabling multi-period electricity market studies, e.g. in the context of flow-based market coupling. 
However, experiments with even larger indicate a need for further study of the design and implementation of the proposed techniques to achieve practical process times.
We reserve these extensions for future work.

\bibliographystyle{IEEEtran}
\bibliography{bibliography.bib}
\appendix
\numberwithin{equation}{subsection}

\subsection{PTDF Derivation}
\label{ax:ptdf_derivation}

Assuming that voltage magnitudes are fixed, phase angle differences between neighbouring nodes are small and reactance dominates resistance on all lines, the active power flow on line $l$ from node $s$ to node $r$ can be written in terms of the phase angle difference between those nodes such that
\begin{align}
    f_{t,l} =  (x_{t,l})^{-1} (\theta_{t,s} - \theta_{t,r}),
    \label{eq:single_line_flow}
\end{align}
where $\theta_{t,n}$ is the voltage angle at node $n$ at time $t$ and we define $\theta_t$ to collect all $\theta_{t,n}, n\in\set{N}$.
All nodal injections and power flows are balanced such that:
\begin{align}
     x_{t,n} = \sum_{\set{L}_n} f_{t,l},
    \label{eq:sinlge_node_powerbalance}
\end{align}
where $\set{L}_n$ is the set of lines connected to node $n$.
Defining incidence matrix ${A\in \{-1,0,1\}^{L\times N}}$ such that all entries are zero except $A_{(l,n)}=1$ if node~$n$ is the sending node of line~$l$ and $A_{(l,n)}=-1$ if $n$ is the receiving node of line~$l$ \eqref{eq:single_line_flow} and \eqref{eq:sinlge_node_powerbalance} can be written in their vector forms as
\begin{align}
    &f_t = X^{-1}A\theta_t = B^{(f)} \theta_t, 
    \label{eq:flows_vector} \\
    &x_t = A^{\!\top} X^{-1} A \theta_t = B^{(n)} \theta_t,
    \label{eq:netinjections_vector}
\end{align}
where diagonal matrix  ${X \in \mathbb{F}^{L\times L}}$ collects line reactances such that $X_{(l,l)} = 1/b_l, \forall l \in \set{L}$ and $B^{(f)}\in \mathbb{R}^{L\times N}$, $B^{(n)}\in \mathbb{R}^{N\times N}$ is the line and bus susceptance matrix, respectively. 
Next, because \eqref{eq:single_line_flow} is based on angle \textit{differences}, we define a reference (slack) node with fixed phase angle.  
Without loss of generality we choose the index of the slack node to be $n_{\text{slack}}=1$. 
Then ${B^0 \in \mathbb{R}^{L\times N}}$ is defined by:
\begin{align}
B^0 = B^{(f)} \begin{bmatrix} 
        0 & 0 \\
        0 & \left(\tilde{B}^{(n)}\right)^{-1}
        \end{bmatrix} \eqqcolon B^{(f)} \hat{B}
        \label{eq:ptdf_formal}
\end{align}
\noindent
where $\tilde{B}^{(n)}\in\mathbb{S}^{N-1}$ is the bus susceptance matrix without the row and column associated with the slack bus (first row and first column in our case).

\subsection{LODF derivation}
\label{ax:lodf_derivation}

Given outage scenario $c$ with $\set{L}_c\subseteq\set{L}$ the set of failed lines,  $\LODF_{l\set{L}_c} \in \mathbb{R}^{1\times|\set{L}_c|}$ can be calculated as \cite[Eq. (3)]{jiachun_guo_direct_2009}:
\begin{equation}
    \LODF_{l\set{L}_c} = (X_{[l]}^{-1} A_l \hat{B} A_{\set{L}_c}^\top)(I - X_{[\set{L}_c]}^{-1} A_{\set{L}_c} \hat{B} A_{\set{L}_c}^\top)^{-1} \label{eq:lodf_definiton}%
\end{equation}%
with $\hat{B}$ as defined in \cref{eq:ptdf_formal} and $X_{[\set{L}_c]}$ the diagonal matrix composed of the rows and columns of $X$ corresponding to the lines in $\set{L}_c$. Thus, if $\set{L}_c$ is a singleton, then $X_{[\set{L}_c]}$ is a scalar. Matrix $A_{\set{L}_c}$ is the $C\times N$ matrix composed of the rows of $A$ corresponding to the lines in $\set{L}_c$. Thus, if $\set{L}_c$ is a singleton, then $A_{\set{L}_c}$ is a row vector. Therefore, $X_{[l]}$ and $A_l$ are a scalar and a row vector, respectively. 

\subsection{Implementation of RayShoot}
\label{ax:rayshoot}

\removelatexerror % lol
\begin{algorithm}[H]
\label{alg:ray_shoot}
\SetAlgoLined
\SetKwFunction{RayShoot}{RayShoot}
\SetKwFunction{Clarkson}{Clarkson}
\SetKwFunction{LPTest}{LPTest}
\SetKwInOut{Input}{input}\SetKwInOut{Output}{output}
\Input{System $(B, \overline{f})$  \\
       Interior point $z$ \\
       Point on or outside of feasible region $x^*$ \\ 
       }
\Output{Index of first inequality that limits a ray starting at $z$ in the direction of $r$}
\Begin{
    $\set{H} \leftarrow \emptyset$; \tcp*[h]{Set of crossed hyperplanes} \\
    $\epsilon \leftarrow \epsilon^{\text{init}}$; \tcp*[h]{Set inital ray increment} \\
    $r = \frac{x^* - z}{\soc{x^* - z}}$; \tcp*[h]{Set direction of ray} \\
    \While{$|\set{H}| \neq 1$}{
        $z \leftarrow z  + \epsilon r$; \tcp*[h]{Add increment to ray} \\
        $\set{H} \leftarrow \{i \mid B_i z > \overline{f}_i \}$\;
        \If{$|\set{H}| > 1 $}{
            $z \leftarrow z  - \epsilon r$; \tcp*[h]{Go back one step} \\
            $\epsilon \leftarrow \epsilon / 10$; \tcp*[h]{Reduce step size} 
        }
    }
    \KwRet{$\set{H}$}
 }
\caption{$\texttt{RayShoot}(B,\overline{f},z,x^*)$}
\end{algorithm}

\vspace{1.5ex}
Note that, because $x^*$ is associated with constraint $k$, see Fig.~\ref{fig:algoillus}, the ray from $z$ to $x*$ will always hit at least $k$ and therefore always return an index.
In practical implementation, if step size $\epsilon$ reaches floating point \textcolor{black}{precision}, \texttt{RayShoot} can return any of the indices of the inequalities that limit the ray from $z$ in the direction $r$.

\end{document}